\documentclass[12pt]{iopart}

\usepackage{mathtext}
\usepackage{amsfonts}
\usepackage{amssymb}

\usepackage{bm}
\usepackage{epsfig}
\usepackage{iopams}

\begin{document}

\title[]{Dynamical behavior of the entanglement, purity and energy between atomic qubits in motion under the influence of thermal environment}

\author{L Tan$^{1,2}$,  Y Q Zhang$^{1}$,  Z H Zhu$^{1}$ and L W Liu$^{1}$}
\address{$^{1}$Institute of Theoretical Physics, Lanzhou
University, Lanzhou 730000, China \\ $^{2}$Key Laboratory for
Magnetism and Magnetic materials of the Ministry of Education,
Lanzhou University, Lanzhou 730000,  China} \ead{Email:
tanlei@lzu.edu.cn}
\def\la{\langle}
\def\ra{\rangle}
\def\om{\omega}
\def\Om{\Omega}
\def\vep{\varepsilon}
\def\wh{\widehat}
\def\tr{\rm{Tr}}
\def\da{\dagger}
\newcommand{\beq}{\begin{equation}}
\newcommand{\eeq}{\end{equation}}
\newcommand{\beqa}{\begin{eqnarray}}
\newcommand{\eeqa}{\end{eqnarray}}
\newcommand{\bV}{{\bf V}}
\newcommand{\bK}{{\bf K}}
\newcommand{\bG}{{\bf G}}
\begin{abstract}
The entanglement, purity and energy of two isolated two-level
atoms which are initially prepared in Bell state and  each
interacts with a thermal cavity field are investigated by
considering  the atomic motion and the field-mode structure. We
achieve the analytical solutions of the atomic qubits by using the
algebraic dynamical approach and the influences of the field-mode
structure parameter, the strength of the thermal field and the
detuning on the entanglement, purity and energy are discussed. We
also investigate the state evolution of the atomic qubits based on
the entanglement-purity-energy diagrams. Our results show that the
disentanglement process of the atomic qubits accompanies by
excitations transferring from atomic subsystem to cavity field
modes and atomic state from a pure state convert to  the mixed
states.
\end{abstract}
\section{Introduction}
Entanglement is one of the most remarkable features of  quantum
mechanics and has many practical applications in  quantum
information processing~\cite{Nielsen00}. However, realistic  quantum
systems are  inevitably influenced by the surrounding environment,
which always  leads to decoherence  of the quantum states.
Particularly, the thermal field is frequently discussed in this
problem. A thermal field, which is emitted by a source in thermal
equilibrium at temperature $T$, is a highly chaotic field with
minimal information about its mean value of the energy. However,
such a chaotic field can entangle qubits that are prepared
initially in a separable state~\cite{Kim02}, lead to  entangled states
in the interaction of a single qubit in a pure state with a
thermal field regardless of the temperature of the field and
reduce the system to a mixed state when the field variables are
traced over~\cite{Bose01}. The influence of the thermal field strength
on the atom-atom entanglement~\cite{Yan09} and atom-field
entanglement~\cite{Yan08} have also been investigated. Besides,
Zheng~\cite{Zheng02} proposed a scheme for realizing two-qubit quantum
phase gates with atoms in a thermal cavity. Jin~\cite{Jin05} suggested
a scheme of  teleporting a two-atom entangled state with a thermal
cavity and the success probability can reach $1.0$.

In this paper, we consider two isolated two-level atoms  each
interacting with a single-mode thermal cavity field. The effects
of the atomic motion and the field-mode structure are considered
at the same time. The atomic motion and the field-mode structure
not only  lead to nonlinear transient effects in the atomic
population~\cite{Schlicher89,Joshi90}, which are similar to self-induced
transparency and adiabatic effects, but also give rise to the
periodic evolution of the entropy squeezing~\cite{Liao04}, the  field
entropy, the atomic inversion~\cite{Fang98} and the
entanglement~\cite{Yan09,Yan08}. Other effects by regulating the
field-mode structure parameters have also been observed, e.g.
decreasing  the squeezing in the two-photon JC model~\cite{Bartzis92},
operating the entanglement and realizing the quantum gate
operation~\cite{Joshi10}. The  recent cavity quantum electrodynamics
experiments  which use an atomic beam passing along the axis of a
rectangular or cylindrical cavity provide the feasibility of
discussing the different  field mode structures~\cite{Meschede85,Rempe87}.
Then, in the present work, we are interested in the  effects  of
the atomic motion and the field-mode structure on the time
evolution of the atomic state in  thermal cavity field
environment. We suppose the two atoms are initially  prepared  in
one of the Bell states, which is a maximal entangled pure state.
The previous studies of maximally entangled states usually use
entanglement, purity and energy  to characterized the set of
two-qubit states~\cite{Cavalcanti06,Ziman05,McHugh06,Yu07}. As the atom-field interacting
process accompanied by  exchanging excitations between atoms and
fields, the energy in the system is a direct influences on the
entanglement and mixedness properties. On the other hand, the
initial pure state of the atomic qubits must be influenced by
interacting with a thermal cavity environment, so the mixedness is
an  important characteristic  in the process of entanglement
evolution. As a result,   we will investigate the time evolution
of the entanglement, purity and energy via manipulating of the
field-mode structure parameters, the strength of the thermal field
and the  detunings between  atoms and  thermal cavity fields. The
relationships among entanglement, pure and energy will also be
presented with entanglement-pure-energy (EPE) diagram, which can
offer a nice visual  to the allowed state of the atomic qubits.

The paper is organized as follows: In Sec.$2$, we first describe
the model under consideration and then derive the exact expression
for the atomic reduced density matrix using algebraic dynamical
approach~\cite{Yu95,Jie97,Xu99,Cen00}. The quantities used to quantify the
entanglement, purity and energy of atomic qubits are also defined
in this section. Sec.$3$ is devoted to investigate  the time
evolution of entanglement, purity and energy for the atomic
qubits. In Sec.$4$, we discuss the time evolution of the atomic
qubits with a EPE diagram. Finally, we present our conclusion in
Sec.$5$.

\section{Model}

We consider  two identical moving  two-level atoms (A and B) and two spatially separated cavities (a and b) with non-decaying single mode  fields
by using  very high quality factor cavities~\cite{Meschede85,Weidinger99}.
Atoms $A$ and $B$  fly through cavities $a$ and $b$ with a constant velocity, respectively.
We suppose the two subsystems $Aa$ and $Bb$ are identical with  same value of  atom-field coupling strength, frequencies
and  field-mode structure.
The Hamiltonian for the  considered system  in the rotating-wave approximation can be written as ($\hbar=1$)
\begin{eqnarray}
H&=&H_{1}+H_{2} ,\nonumber\\
H_{1}&=&\omega_{c}a^{\dagger}a+\omega_{0}S_{z}^{A}+gf(z)(a^{\dagger}S_{-}^{A}+aS_{+}^{A}),\nonumber\\
H_{2}&=&\omega_{c}b^{\dagger}b+\omega_{0}S_{z}^{B}+gf(z)(b^{\dagger}S_{-}^{B}+bS_{+}^{B})
\end{eqnarray}
where $H_{1}$ and $H_{2}$ are the Hamiltonians   for subsystems Aa and Bb, respectively.
$a^{\dagger}$ and $a$  ($b^{\dagger}$ and $b$ ) are the creation and annihilation
operators of the cavity field a (b). $S_{+}^{i}$, $S_{-}^{i}$ and $S_{z}^{i}$ represent the
atomic raising, lowering and inversion operators of the atom $i(i=A,B)$. $\omega_{c}$ and $\omega_{0}$ are the frequencies for
the field $a$ and the atom $A$ (or the field $b$ and the atom $B$), respectively. $g$ is the atom-field coupling
strength, and $f(z)$ is the shape function of the cavity field mode.
When the interaction energy of atom-field coupling is much larger
than the transverse kinetic energy spread of the atom, we can neglect
the transverse  velocity spread  and  restrict our investigation to atomic motion along the cavity axis ($z$ axis).
Then the atomic motion is incorporated into $f(z)$ as follows
\begin{eqnarray}
f(z)\rightarrow f(\upsilon t),
\end{eqnarray}
where $\upsilon$ is the atomic motion velocity. In this regard  the
cavity field-mode $TEM_{mnp}$ is defined like
$f(\upsilon t)=\sin(p\pi\upsilon t/L)$,
where $p$ represents the number of half wavelengths of the
field-mode inside a cavity with length $L$.
If the  atom  passes through the cavity  so fast that
the atomic motion can be considered as a constant.
For a proper choice of the atomic motion velocity $\upsilon=gL/\pi$,
then $\int_{0}^{t}f(\upsilon
t^{'})dt^{'}=[1-\cos(pgt)]/pg$.

In the following, we propose the algebraic dynamical approach to derive the time evolution
operator and the density operator based on  the Hamiltonian (1).
The key idea of the algebraic dynamical approach is introducing a canonical transformation operator
that transforms the Hamiltonian into a liner function in terms of a set of Lie algebraic generators.
According to algebraic dynamics, linear systems are integrable and solvable, then the time
evolution operator and the density operator can be obtained easily.
In the case of symmetric
atom-field interaction, the two subsystems are completely equivalent. For simplicity,
we will work with the subsystem $Aa$.  A straightforward analysis of the Hamiltonian (1) shows
that the total excitation number for  subsystem  $Aa$ is
\begin{eqnarray}
N_{1}=a^{\dagger}a+S_{z}^{A}+\frac{1}{2},
\end{eqnarray}
which is a conserved quantity for the subsystem Aa and commutes with the Hamiltonian $H_{1}$. Based on the algebraic
dynamical approach, introducing  SU(2) algebra generators $\{J_{0}$,
$J_{+}$, $J_{-}\}$, with $J_{0}=S_{z}^{A}$, $J_{+}=N_{1}^{-1/2}aS_{+}^{A}$,
$J_{-}=N_{1}^{-1/2}a^{\dagger}S_{-}^{A}$, which are nonlinear expressions
and obey the following commutation relations
\begin{eqnarray}
[J_{0},J_{+}]=J_{+},\quad [J_{0},J_{-}]=-J_{-}, \quad
[J_{+},J_{-}]=2J_{0}.
\end{eqnarray}

In terms of the SU(2) algebra generators and the canonical
transformation operator $U_{g}=\exp(\theta J_{+}-\theta J_{-})$, we can obtained the time
evolution operator of the subsystem Aa
\begin{eqnarray}
U_{1}(t)&=&e^{-iH_{1}dt}=U_{g}e^{-i(U_{g}^{-1}H_{1}U_{g})t }U_{g}^{-1}\nonumber\\
&=& e^{-iE_{1}t}[\cos\frac{\lambda t}{2}-2i J_{0}\sin\frac{\lambda
t}{2}\cos 2\theta + i(J_{+}+J_{-})\sin\frac{\lambda
t}{2}\sin2\theta].
\end{eqnarray}
where $E_{1}=\omega_{c}(N_{1}-\frac{1}{2})$, $\theta=-\arctan[(\sqrt{\Delta^{2}/4+g'^{2}N_{1}}-\Delta/2)/g'N_{1}^{1/2}]$,
$\lambda=\sqrt{\Delta^{2}+4g'^{2}N_{1}}$, $g'=g\alpha/t$, $\alpha=\int_{0}^{t}f(\upsilon
t^{'})dt^{'}=[1-\cos(pgt)]/pg$ and
$\Delta=\omega_{0}-\omega_{c}$ is the detuning between the atom $A$ and the cavity $a$.

What should be noticed  here is that using canonical transformation
operator to diagonalize the nonlinear Hamiltonian (1) doesn't change
its intrinsic qualities. Likewise, we can get the evolution operator $U_{2}$ of the  subsystem Bb, which has the similar form as $U_{1}$.
\begin{eqnarray}
U_{2}(t)=e^{-iE_{2}t}[\cos\frac{\eta t}{2}-2i L_{0}\sin\frac{\eta t}{2}\cos 2\phi + i(L_{+}+L_{-})\sin\frac{\eta t}{2}\sin2\phi].
\end{eqnarray}
where $E_{2}=\omega_{c}(N_{2}-\frac{1}{2})$, $\phi=-\arctan[(\sqrt{\Delta^{2}/4+g'^{2}N_{2}}-\Delta/2)/g'N_{2}^{1/2}]$,
and $\eta=\sqrt{\Delta^{2}+4g'^{2}N_{2}}$. $N_{2}=b^{\dagger}b+S_{z}^{B}+\frac{1}{2}$
is the total excitation number for  subsystem  $Bb$.
$\{L_{0}$, $L_{+}$, $L_{-}\}$ are the  SU(2) algebra generators with $L_{0}=S_{z}^{B}$, $L_{+}=N_{2}^{-1/2}bS_{+}^{B}$,
$L_{-}=N_{2}^{-1/2}b^{\dagger}S_{-}^{B}$.

Throughout this paper we suppose the two atoms $AB$ to be initially prepared in one of the
Bell states, $|\Psi\rangle=\frac{1}{\sqrt{2}}(|eg\rangle+|ge\rangle)$, and the two thermal cavity fields $ab$
are in single-mode thermal field states $\rho_{a}(0)=\sum_{n=0}^{\infty}P_{n}|n\rangle\langle n|$,
$\rho_{b}(0)=\sum_{m=0}^{\infty}P_{m}|m\rangle\langle m|$.  As a result, the initial density operators
for the two atoms and the two thermal cavity fields can be written as
\begin{eqnarray}
\rho_{AB}(0)&=&|\Psi\rangle\langle\Psi|=\frac{1}{2}(|eg\rangle\langle eg|+|eg\rangle\langle ge|+|ge\rangle\langle eg|+|ge\rangle\langle ge|),\nonumber\\
\rho_{f}(0)&=&\rho_{a}(0)\otimes\rho_{b}(0)=\sum_{n=0}^{\infty}\sum_{n=0}^{\infty}P_{n}P_{m}|nm\rangle\langle nm|,
\end{eqnarray}
where $P_{n}=\frac{k^{n}}{(k+1)^{n+1}}$,$P_{m}=\frac{l^{m}}{(l+1)^{m+1}}$.
$k=1/[\exp(\omega_{c}/T_{a})-1]$ and $l=1/[\exp(\omega_{c}/T_{b})-1]$, $k$ and $l$
are the mean photon numbers of the thermal cavity field mode a and  the thermal  cavity field mode b,
corresponding to the  temperatures $T_{a}$ and $T_{b}$,   respectively.

Then, the initial density operator for the total system can be derived as
\begin{eqnarray}
\rho_{AB-f}(0)&=& \rho_{AB}(0)\otimes\rho_{f}(0)\nonumber\\
&=&\frac{1}{2}\sum_{n}\sum_{m}P_{n}P_{m}(|engm\rangle\langle engm|+|engm\rangle\langle gnem|\nonumber\\
&+&|gnem\rangle\langle engm|+|gnem\rangle\langle gnem|),
\end{eqnarray}
where the $|engm\rangle$  indicates that atom $A$ is in the excited state and atom  $B$ is in the ground state,
field mode $A$ and field mode $B$ are in  the states $ |n\rangle$ and $|m\rangle$, respectively.

The initial state (8) under the action of the operator $U_{1}(t)\otimes U_{2}(t)$ evolves to
\begin{eqnarray}
\rho_{AB-f}(t)=U_{1}(t)U_{2}(t)\rho_{AB-f}(0)U_{2}^{\dagger}(t)U_{1}^{\dagger}(t).
\end{eqnarray}

Then, from Eq.(9), we can get the reduced density matrix $\rho_{AB}(t)$
of the subsystem $AB$  by tracing over the thermal cavity field variables.
In terms of the atomic
basis states $|gg\rangle$, $|ge\rangle$, $|eg\rangle$, and $|ee\rangle$, the reduced density operator $\rho_{AB}(t)$ can be expressed as
\begin{equation}
\rho_{AB}(t)=Tr_{f}[\rho_{AB-f}(t)]=\left(
\begin{array}{cccc}
 x_{1} & 0 & 0 & 0 \\
 0 & x_{2} & x_{3} & 0 \\
 0 &x_{4} & x_{5} & 0 \\
 0 & 0 & 0 &  x_{6}
\end{array}
\right)
\end{equation}
where $x_{1}+x_{2}+x_{5}+x_{6}=1$,
\begin{eqnarray}
x_{1}&=&\frac{1}{2}\sum_{n}\sum_{m}\{P_{n-1}P_{m}
[\sin^{2}(\frac{\lambda_{n}t}{2})\sin^{2}(2\theta_{n})][\cos^{2}(\frac{\eta_{m}t}{2})+\sin^{2}(\frac{\eta_{m}t}{2})\cos^{2}(2\phi_{m})]\nonumber\\
&+&P_{n}P_{m-1}[\cos^{2}(\frac{\lambda_{n}t}{2})+\sin^{2}(\frac{\lambda_{n}t}{2})\cos^{2}(2\theta_{n})]
[\sin^{2}(\frac{\eta_{m}t}{2})\sin^{2}(2\phi_{m})]\},\nonumber\\
x_{2}&=&\frac{1}{2}\sum_{n}\sum_{m}\{P_{n}P_{m}
[\cos^{2}(\frac{\lambda_{n}t}{2})+\sin^{2}(\frac{\lambda_{n}t}{2})\cos^{2}(2\theta_{n})]\nonumber\\
&\times&[\cos^{2}(\frac{\eta_{m+1}t}{2})+\sin^{2}(\frac{\eta_{m+1}t}{2})\cos^{2}(2\phi_{m+1})]\nonumber\\
&+&P_{n-1}P_{m+1}
[\sin^{2}(\frac{\lambda_{n}t}{2})\sin^{2}(2\theta_{n})][\sin^{2}(\frac{\eta_{m+1}t}{2})\sin^{2}(2\phi_{m+1})]\},\nonumber\\
x_{3}&=&\frac{1}{2}\sum_{n}\sum_{m}\{P_{n}P_{m}
[\cos(\frac{\lambda_{n}t}{2})+i\sin(\frac{\lambda_{n}t}{2})\cos2\theta_{n})]\nonumber\\
&\times&[\cos(\frac{\lambda_{n+1}t}{2})+i\sin(\frac{\lambda_{n+1}t}{2})\cos(2\theta_{n+1})]\nonumber\\
&\times&[\cos(\frac{\eta_{m+1}t}{2})-i\sin(\frac{\eta_{m+1}t}{2}\cos2\phi_{m+1})][\cos(\frac{\eta_{m}t}{2})-i\sin(\frac{\eta_{m}t}{2})\cos(2\phi_{m})]\},
\nonumber\\
x_{4}&=&x_{3}^{\ast},\nonumber\\
x_{5}&=& \frac{1}{2}\sum_{n}\sum_{m}\{P_{n+1}P_{m-1}
[\sin^{2}(\frac{\lambda_{n+1}t}{2})\sin^{2}(2\theta_{n+1})][\sin^{2}(\frac{\eta_{m}t}{2})\sin^{2}(2\phi_{m})]\nonumber\\
&+&P_{n}P_{m}
[\cos^{2}(\frac{\lambda_{n+1}t}{2})+\sin^{2}(\frac{\lambda_{n+1}t}{2})\cos^{2}(2\phi_{n+1})]\nonumber\\
&\times&[\cos^{2}(\frac{\eta_{m}t}{2})+\sin^{2}(\frac{\eta_{m}t}{2})\cos^{2}(2\phi_{m}]\},\nonumber\\
x_{6}&=&\frac{1}{2}\sum_{n}\sum_{m}\{P_{n}P_{m+1}
[\cos^{2}(\frac{\lambda_{n+1}t}{2})+\sin^{2}(\frac{\lambda_{n+1}t}{2})\cos^{2}(2\theta_{n+1})]\nonumber\\
&\times&[\sin^{2}(\frac{\eta_{m+1}t}{2})\sin^{2}(2\phi_{m+1})]\nonumber\\
&+&P_{n+1}P_{m}
[\sin^{2}(\frac{\lambda_{n+1}t}{2})\sin^{2}(2\theta_{n+1})][\cos^{2}(\frac{\eta_{m+1}t}{2})+\sin^{2}(\frac{\eta_{m+1}t}{2})\cos^{2}(2\phi_{m+1})]\},\nonumber\\
\end{eqnarray}%
and $\lambda_{n}=\sqrt{\Delta^{2}+4g'^{2}n}$, $\eta_{m}=\sqrt{\Delta^{2}+4g'^{2}m}$,
$\theta_{n}=-\arctan[(\sqrt{\Delta^{2}/4+g'^{2}n}-\Delta/2)/g'n^{1/2}]$,
$\phi_{m}=-\arctan[(\sqrt{\Delta^{2}/4+g'^{2}m}-\Delta/2)/g'm^{1/2}]$.

Based on the analytical solution of $\rho_{AB}(t)$, we can conveniently do some approximation
in analyzing the numerical results in the following section.
Besides, we can define the quantities to quantify the entanglement, purity and energy
of the atomic qubits and give their expressions in terms of the matrix elements of $\rho_{AB}(t)$.

We adopt Wootters' concurrence as a measure of entanglement in this discussion~\cite{Wootters98}, which is denoted as
$C_{AB}=Max\{0,\sqrt{\lambda_{1}}-\sqrt{\lambda_{2}}-\sqrt{\lambda_{3}}-\sqrt{\lambda_{4}}\}$,
and $\lambda_{i}$ are the eigenvalues of  the matrix $(\rho_{AB}\tilde{\rho}_{AB})$ in non increasing order.
Following Eq.(10), the expression  of  $C_{AB}$ turns out to be \begin{eqnarray}
C_{AB}=2Max\{0,|x_{3}|-\sqrt{x_{1}\times x_{6}}\}.
\end{eqnarray}

When the value of  $C_{AB}$ is positive,  the atomic system is entangled.
$C_{AB}=1$ corresponds to  the  maximal entanglement state, while $C_{AB}=0$ indicates the atom A and the atom B are separable.

The energy $U_{AB}$ of subsystem $AB$ is defined here as the expectation
value of the Hamiltonian $H_{AB}=\omega_{0}S_{z}^{A}+\omega_{0}S_{z}^{B}$.
We set $\omega_{0}=1$, then the energy $U_{AB}$ can be obtained  based on the expression of Eq.(10)
\begin{eqnarray}
U_{AB}=Tr\{\rho_{AB-f}(t)H_{AB}\}=x_{6}-x_{1}
\end{eqnarray}

For two two-level atoms, $U_{AB}$ ranges from -1 for $\rho_{AB}=|gg\rangle\langle gg|$ to 1 for $\rho_{AB}=|ee\rangle\langle ee|$.

To quantify the mixedness of the state $\rho_{AB}(t)$, we use the purity
\begin{eqnarray}
P_{AB}=Tr\{\rho_{AB}^{2}(t)\}=x_{1}^{2}+ x_{2}^{2}+ x_{5}^{2}+ x_{6}^{2}+2x_{3}x_{4}
\end{eqnarray}

For the atomic qubits, $P_{AB}$ ranges from 1/d for completely mixed state to 1 for pure state for d-dimensional systems,
which is closely to the linear entropy measure of mixedness.

The aim of this paper is to address the question how the atomic motion and
the field-mode structure  influence the state  $\rho_{AB}(t)$ of the atomic qubits  in the cases of thermal environment.
We know that the two atoms are initially in  the maximal entanglement state,
and the atomic state will evolves with time  followed by the variation of the entanglement, purity
and the transfer of the energy.
Their time evolution will be discussed in the next section.
\section{Entanglement, purity and energy versus time}
There are three controllable parameters  in the analytical expression of $\rho_{AB}(t)$: the field-mode structure parameter,
the mean photon number in each cavity and the detuning. In this section,
we will discuss their effects of the three parameters on the time evolution of the entanglement, purity and energy of the atomic subsystem.

In Fig.1 and  Fig.2 we plot the time evolution of  $C_{AB}$,  $P_{AB}$ and  $U_{AB}$ affected by different values of
the field-mode structure parameters and the mean photon number in
the situation of exact resonance.
Form Eq.(11), we can easily find that
for resonant atom-field coupling, $\Delta$=0,
$\lambda_{n}=2g'\sqrt{n}$, $\eta_{m}=2g'\sqrt{m}$, $\sin2\theta_{n}=\sin2\phi_{m}=-1$,
$\cos2\theta_{n}=\cos2\phi_{m}=0$. The elements of the matrix $\rho_{AB}$ expressed in Eq.(11)
convert to

\begin{eqnarray}
x_{1}&=&\frac{1}{2}\sum_{n}\sum_{m}[P_{n-1}P_{m}
\sin^{2}(g't\sqrt{n})\cos^{2}(g't\sqrt{m})\nonumber\\
&+&P_{n}P_{m-1}
\cos^{2}(g't\sqrt{n})\sin^{2}(g't\sqrt{m})],\nonumber\\
x_{2}&=&\frac{1}{2}\sum_{n}\sum_{m}[P_{n}P_{m}
\cos^{2}(g't\sqrt{n})\cos^{2}(g't\sqrt{m+1})\nonumber\\
&+&P_{n-1}P_{m+1}
\sin^{2}(g't\sqrt{n})\sin^{2}(g't\sqrt{m+1})],\nonumber\\
x_{3}&=&\frac{1}{2}\sum_{n}\sum_{m}[P_{n}P_{m}\cos(g't\sqrt{n})\cos(g't\sqrt{n+1})
\cos(g't\sqrt{m})\cos(g't\sqrt{m+1})],\nonumber\\
x_{4}&=&x_{3},\nonumber\\
x_{5}&=& \frac{1}{2}\sum_{n}\sum_{m}[P_{n+1}P_{m-1}
\sin^{2}(g't\sqrt{n+1})\sin^{2}(g't\sqrt{m})\nonumber\\
&+&P_{n}P_{m}\cos^{2}(g't\sqrt{n+1})
\cos^{2}(g't\sqrt{m})],\nonumber\\
x_{6}&=&\frac{1}{2}\sum_{n}\sum_{m}[P_{n}P_{m+1}
\cos^{2}(g't\sqrt{n+1})\sin^{2}(g't\sqrt{m+1})\nonumber\\
&+&P_{n+1}P_{m}
\sin^{2}(g't\sqrt{n+1})\cos^{2}(g't\sqrt{m+1})].
\end{eqnarray}
%%
%-------------------------------------------------------------------
\begin{figure}
\begin{center}
\includegraphics[height=3cm,angle=0]{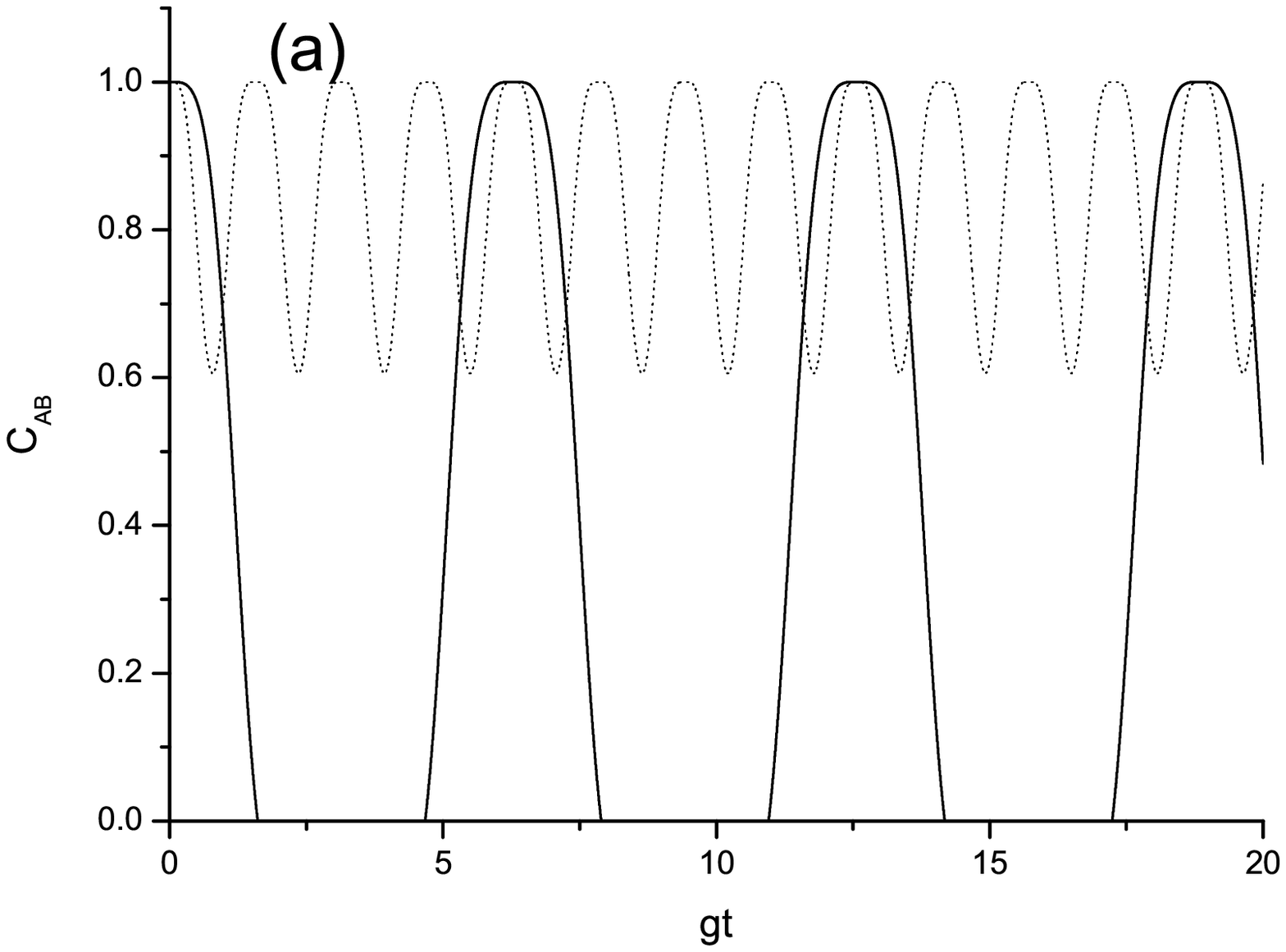}
\includegraphics[height=3cm,angle=0]{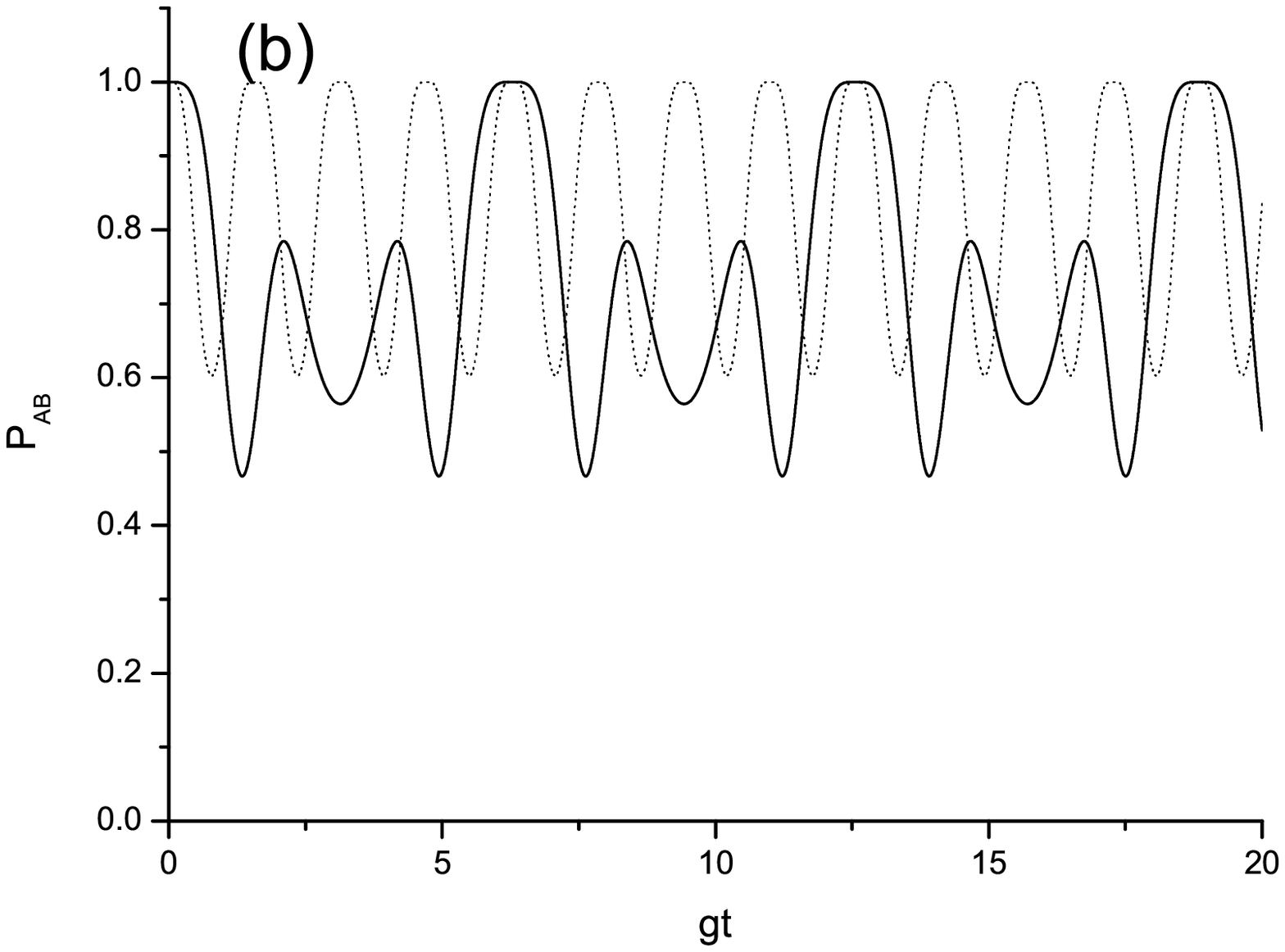}
\includegraphics[height=3cm,angle=0]{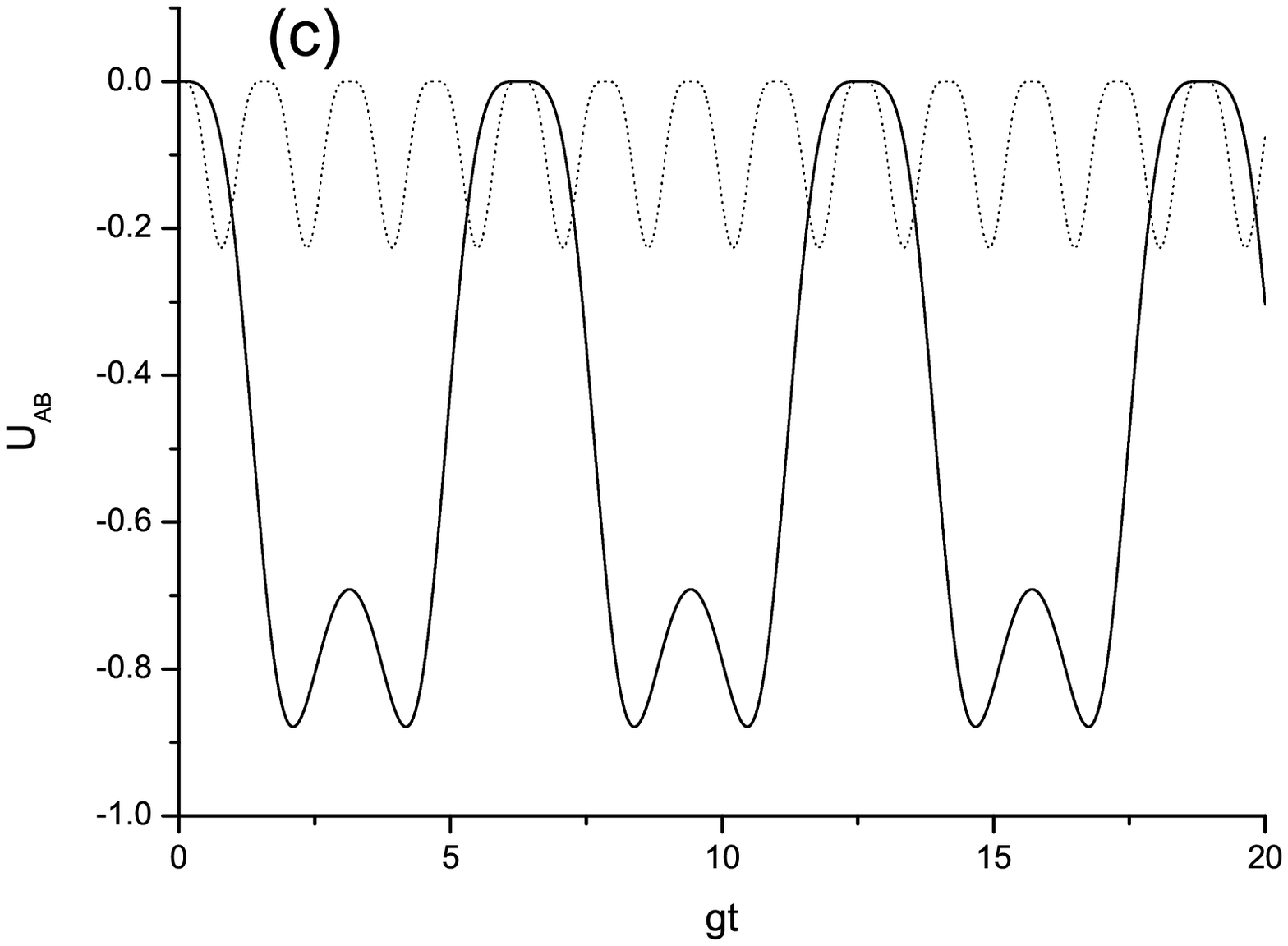}
\vspace*{.2cm}
\caption{\label{apop}
The evolution of  $C_{AB}$,  $P_{AB}$ and  $U_{AB}$ for different field-mode structure parameters, $p=1$(solid)
and $p=4$(dotted). The two subsystems are symmetric for $k=l=0.1$ and the atom and the field are in  exact resonance.
(a) $C_{AB}$ versus time $gt$;  (b) $P_{AB}$ versus time $gt$; (c)  $U_{AB}$ versus time $gt$.
%The corresponding pulses of $^{87}$Rb atoms are plotted for
%$t=20$ms with $\tau=0.5$ms.
}
\end{center}
\end{figure}
%----------------------------------------------------------------------
%
%Fig. \ref{apop} shows the details of the probability density for a single pulse
%and different aperture functions, where all states are
%normalised to one in coordinate space.
%As a result of smoothing the time window function, the sidelobes are highly suppressed.
%This is a direct consequence of the broader energy distribution induced by apodization.
%
%-------------------------------------------------------------------
%
\subsection{the effects of field-mode structure parameters}
The  atomic initial state is a  Bell state, which is a pure state with  $P_{AB}=1$ as well as a maximal entanglement state with $C_{AB}=1$.
However, the initial energy for the atomic subsystem is zero, that is  $U_{AB}=0$.
Fig.1 illustrates the cases when the atom is in
motion at the velocity $\upsilon=gL/\pi$ for parameters $p=1$ and $p=4$, respectively.
It has   been studied  that when the atomic motion is considered, the time
behaviors of field entropy, atomic inversion~\cite{Fang98}, and entropy squeezing~\cite{Liao04} are  periodical, and
their evolution periods are shorten with the increase of  parameter $p$. Similar behaviors occur
in this work. From  Fig.1  we can find that  the evolution periods  is
decreased  with the increase of parameter $p$. This is because the time factor is the scaled time $gt$ when
the atomic motion is neglected, and is $g't$ when the atomic motion is taken into
account. $g't=[1-\cos(pgt)]/p$ is a periodical function on the scaled time $gt$ with period $2\pi/p$.
In addition, the amplitudes for $C_{AB}$, $P_{AB}$ and $U_{AB}$
are reduced while their maximum values are still unchanged.
That is, compared with $p=1$, the  ``sudden death of entanglement" disappears,
the maximum mixedness of the atomic state  reduces
and the energy exchange  between atoms and field modes decreases
with the increase of the field-mode structure parameter $p$.
\subsection{the effects of mean photon number in each cavity}
What we  talked about in Fig.1 is just  limited to the situation of very weak thermal field with mean photon number $k=l=0.1$ in each cavities.
Then we are interested in how the atomic qubits evolves
as the  mean photon number in each cavity increases.
It has been demonstrated that thermal cavity field can lead to  entangled state of quantum qubits interacting with it~\cite{Kim02,Bose01}, while strong
thermal cavity field can inhibit the atom-atom entanglement~\cite{Yan09} and atom-field entanglement~\cite{Yan08}.
Here, we pay our attention to the influence of  mean photon number on the non-local atom-atom entanglement,
in the  case of exact resonance, as is shown in Fig.2(a).
In addition, in Fig.2(b) and  Fig.2(c), further study is employed to the time evolution of  purity and energy,  which can help us to get more
information about entanglement evolution.
From  Fig.2 we can find that with the increase of  mean photon number in each cavity, both
the amplitudes and the maximum values for  $C_{AB}$, $P_{AB}$ and $U_{AB}$ decrease.
That is, compared with the case of  weak thermal  cavity field,
the atomic qubits, which couples  to  strong thermal cavity fields,
can not evolve to a maximal entanglement state with $C_{AB}=1$ as well as a pure state with $P_{AB}=1$.
Meanwhile, the time interval of  the ``sudden death of entanglement" lengthens, the maximum mixedness  increases,
while the energy transfer between atomic qubits and field modes in each interacting period is more and more less.
%%
%-------------------------------------------------------------------
\begin{figure}
\begin{center}
\includegraphics[height=3cm,angle=0]{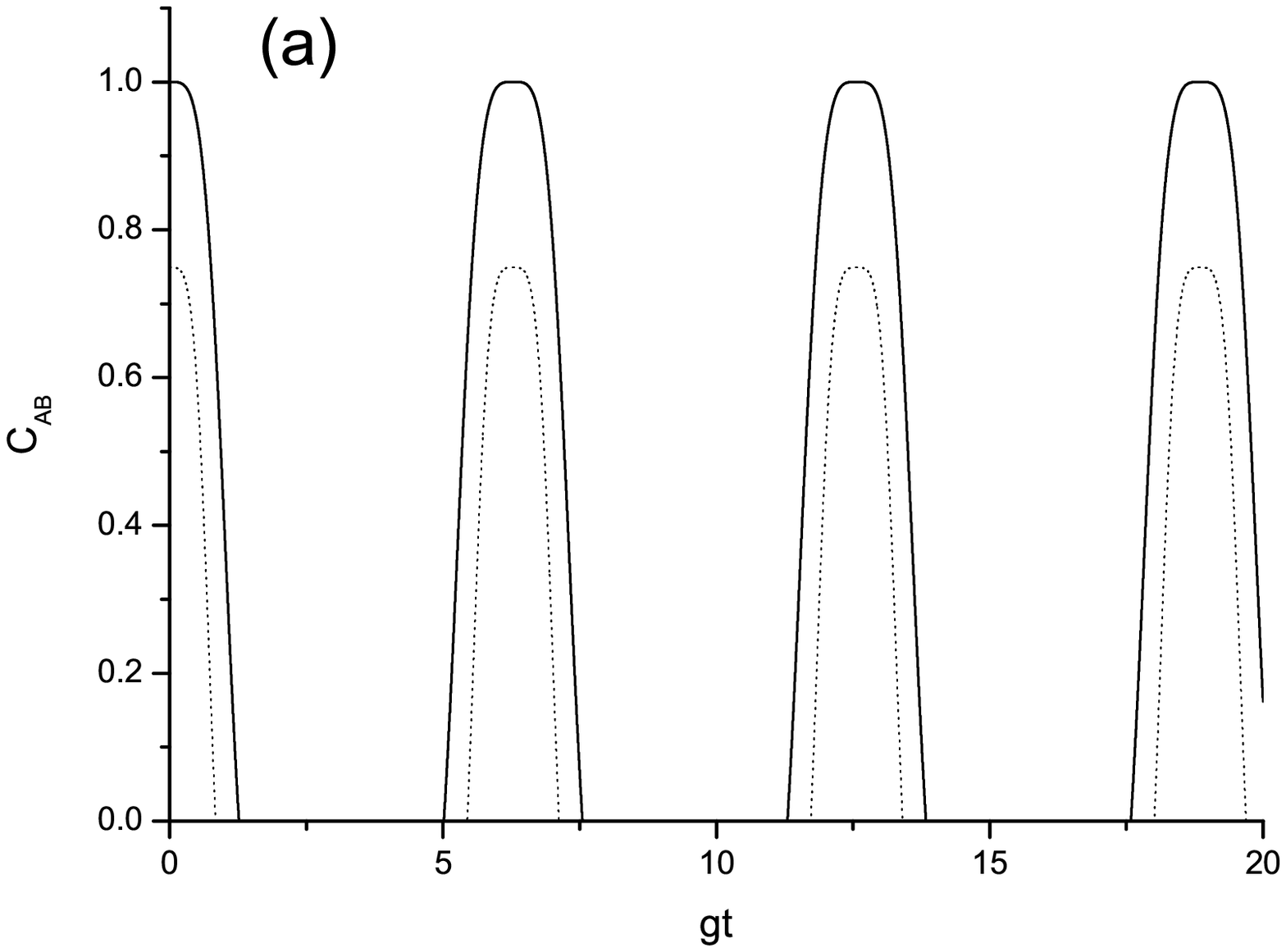}
\includegraphics[height=3cm,angle=0]{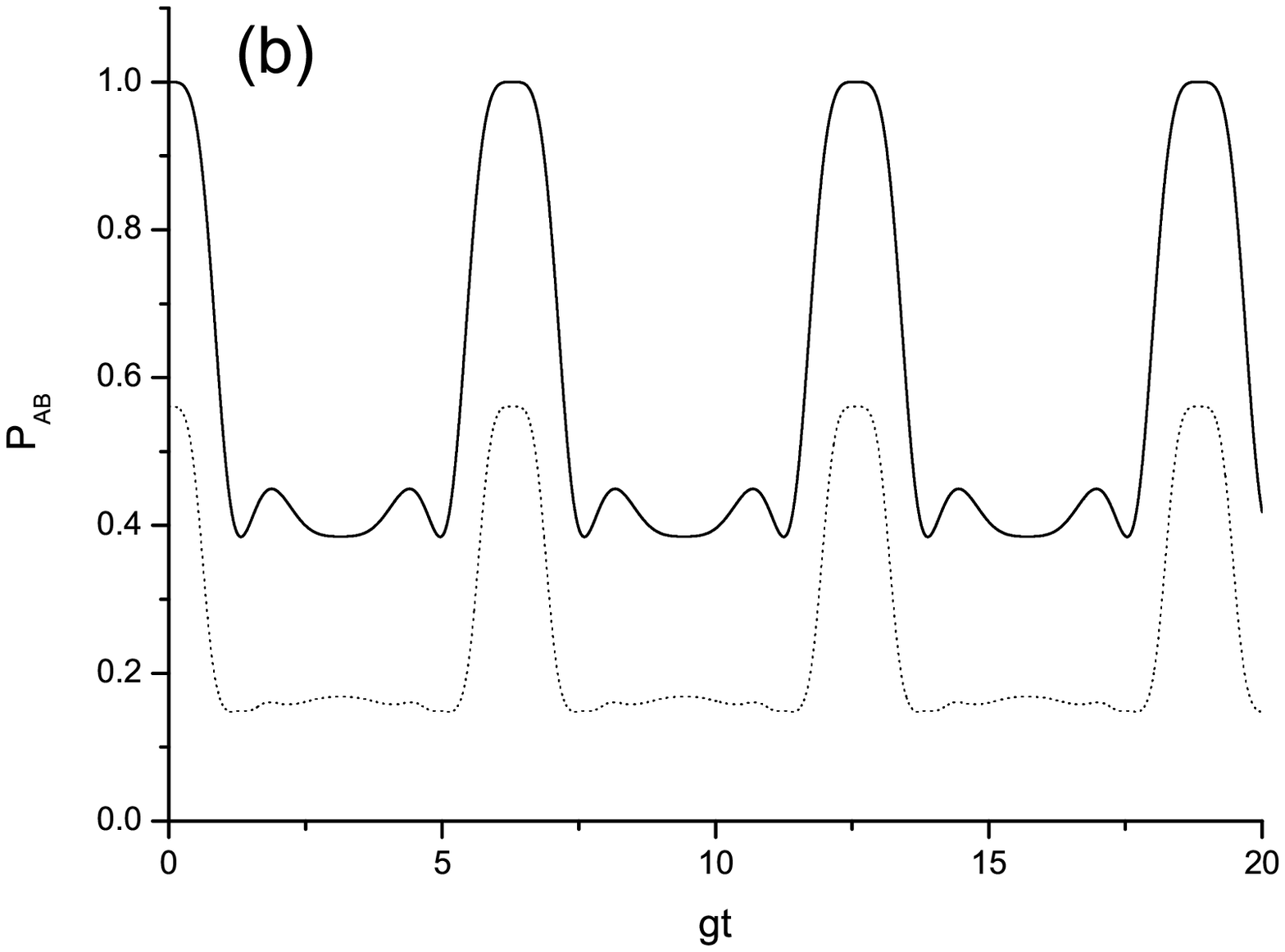}
\includegraphics[height=3cm,angle=0]{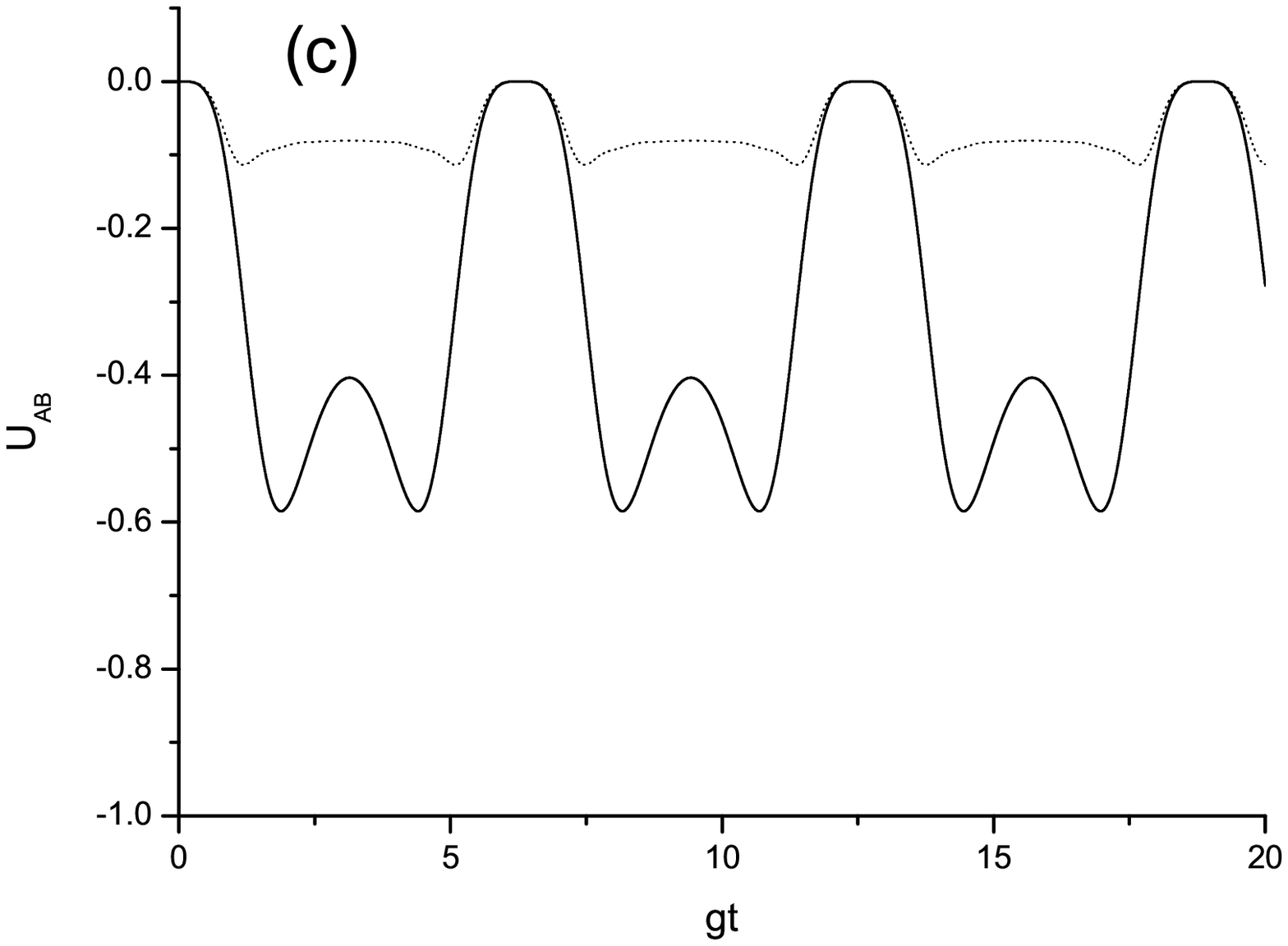}
\vspace*{.2cm}
\caption{\label{xpulses}
The evolution of  $C_{AB}$,  $P_{AB}$ and  $U_{AB}$ for different thermal field strengths, $k=l=0.5$(solid),
$k=l=5$(dotted). The atom and the field are in  exact resonance and the field-mode structure parameter $p=1$.
(a) $C_{AB}$ versus time $gt$;  (b) $P_{AB}$ versus time $gt$; (c)  $U_{AB}$ versus time $gt$.}
\end{center}
\end{figure}
%--------%
%-------------------------------------------------------------------
%--------
%
%
\subsection{the effects of detunings between atom and field}
The effects of detuning on the entanglement, purity and energy between atom A and atom B are depicted in  Fig.3.
In   Fig.3(a), $C_{AB}$  oscillates at first and as time evolves
it will maintain the maximal entanglement state when  the influence of the atomic motion is considered and $\Delta\neq0$.
Moreover, the larger the value of detuning, the faster the  $C_{AB}$ reaches the stable maximal value 1.
When  $\Delta\gg g$, the oscillating almost vanishes and the two atoms  nearly entangle  maximally  all the time as  time evolves.
Similar behavior to the time evolution of $P_{AB}$  is depicted in Fig.3(b). In Fig.3(c)  we can find that with the increase of $\Delta$, the amplitude of $U_{AB}$ is reduced gradually, and large values of $\Delta$ make it more easy to get to zero.
In a word, with the increase of detuning, the atomic subsystem is almost "frozen" in the initial state.
This can be explained as follows:
on the one hand, based on  the expression of $\rho_{AB}$,
for weak thermal cavity fields $k=l=0.1$ and large detuning $\Delta\gg g$, $\lambda_{n}=\eta_{m}\approx\Delta$,  $\sin2\theta_{n}=\sin2\phi_{m}\approx0$,
$\cos2\theta_{n}=\cos2\phi_{m}\approx1$. As a result, $x_{1}=x_{6}\approx0$,
$x_{2}=x_{3}=x_{4}=x_{5}\approx\frac{1}{2}\sum_{n}\sum_{m}P_{n}P_{m}$.
Then, $C_{AB}=2Max\{0,|x_{3}|-\sqrt{x_{1}\times x_{6}}\}\approx\sum_{n}\sum_{m}P_{n}P_{m}=1$,
$P_{AB}=Tr\{\rho_{AB}^{2}(t)\}=x_{1}^{2}+ x_{2}^{2}+ x_{5}^{2}+ x_{6}^{2}+2x_{3}x_{4}\approx\sum_{n}\sum_{m}P_{n}P_{m}=1$,
$U_{AB}=Tr\{\rho_{AB-f}(t)H_{AB}\}=x_{6}-x_{1}\approx0$;
on the other hand, the interacting process between atoms and fields is
accompanied by the transfer of the excitation between the localized atom and cavity mode,
which depends on the atom-field coupling and is distinctly influenced by the value of detuning.
Larger detuning can inhibit the atom-field coupling and restrain this transfer process greatly,
therefore the initially maximal entanglement pure state can be "frozen" in the atomic subsystem.
%
%-------------------------------------------------------------------
\begin{figure}
\begin{center}
\includegraphics[height=3cm,angle=0]{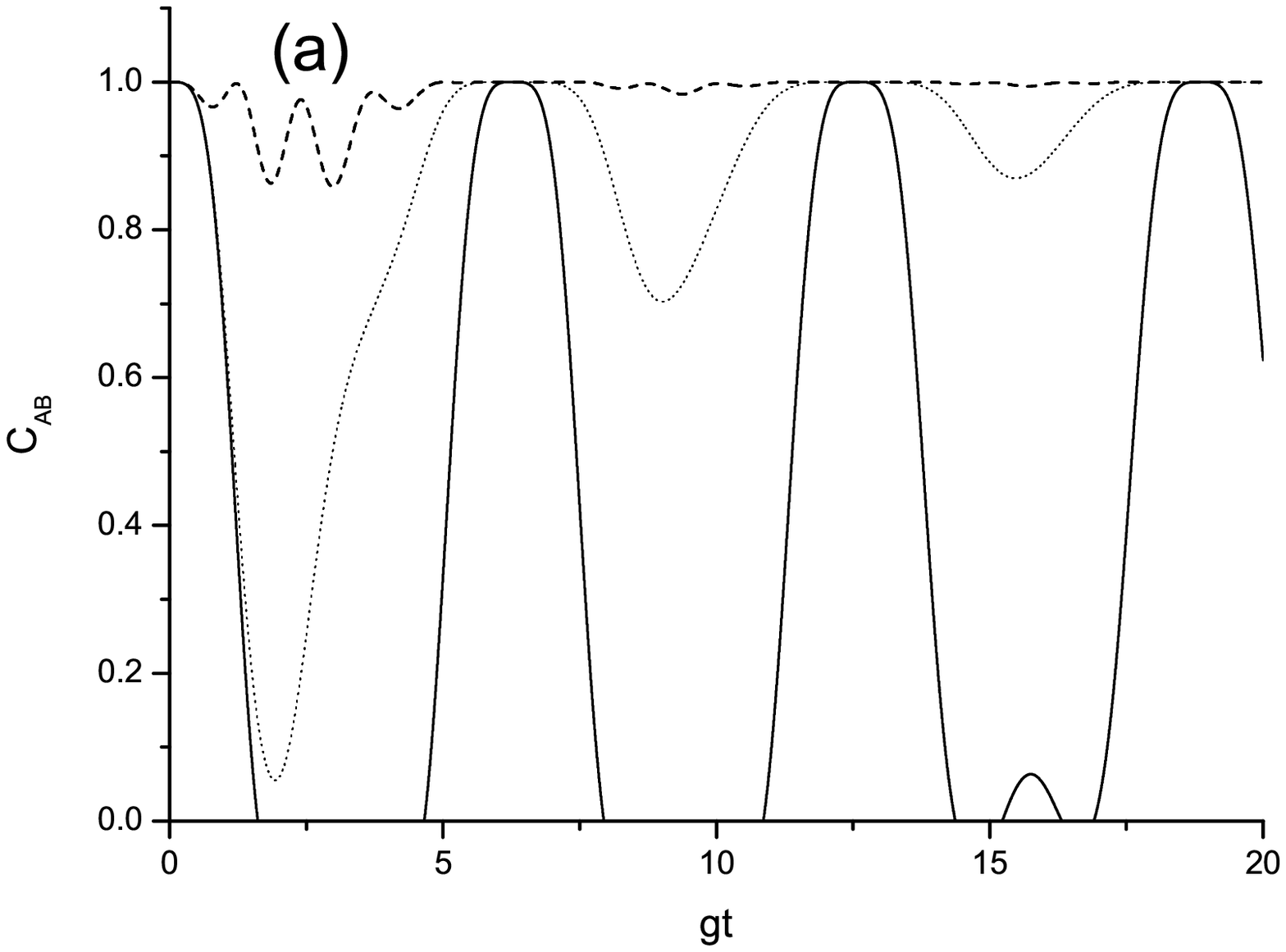}
\includegraphics[height=3cm,angle=0]{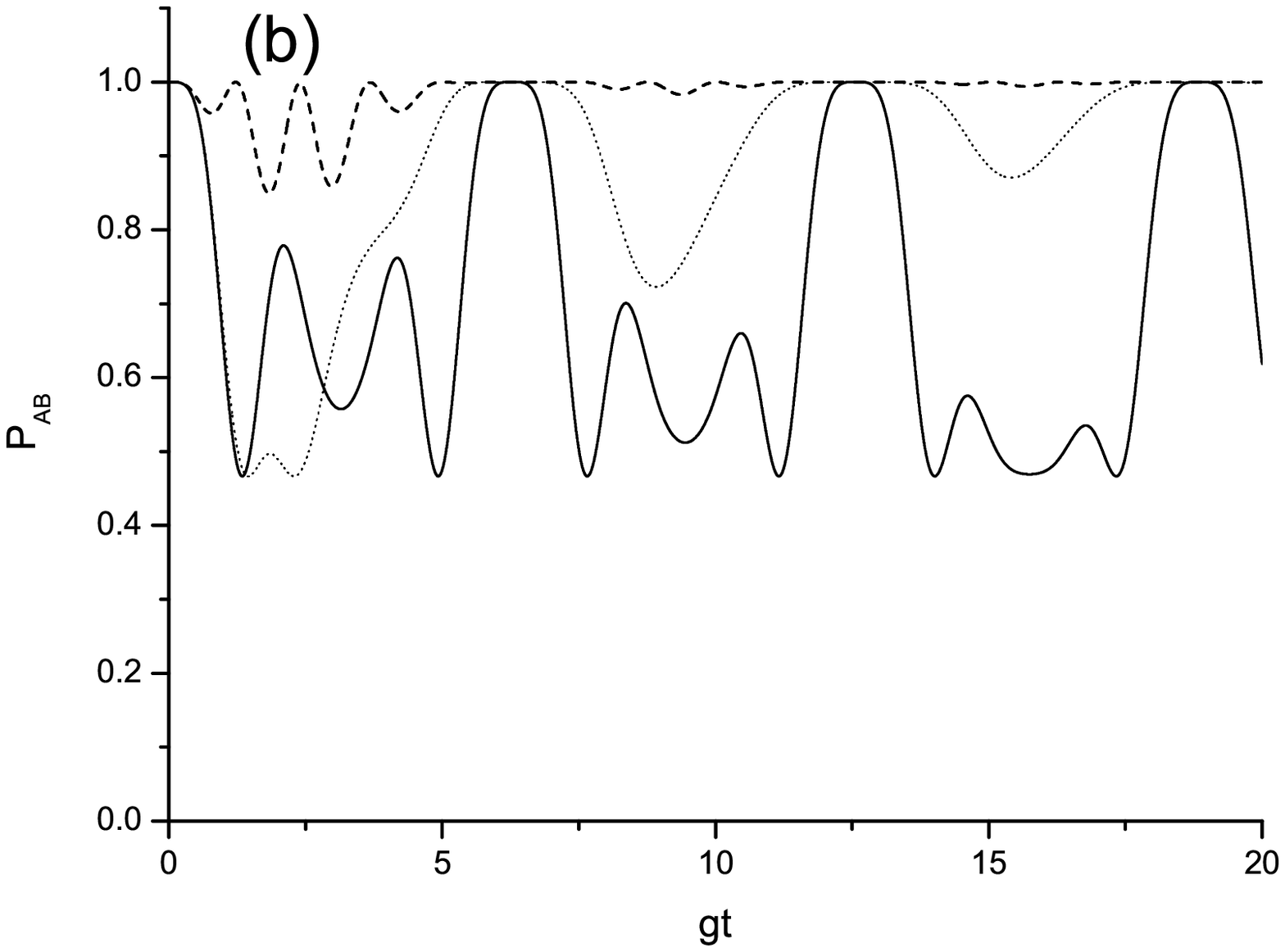}
\includegraphics[height=3cm,angle=0]{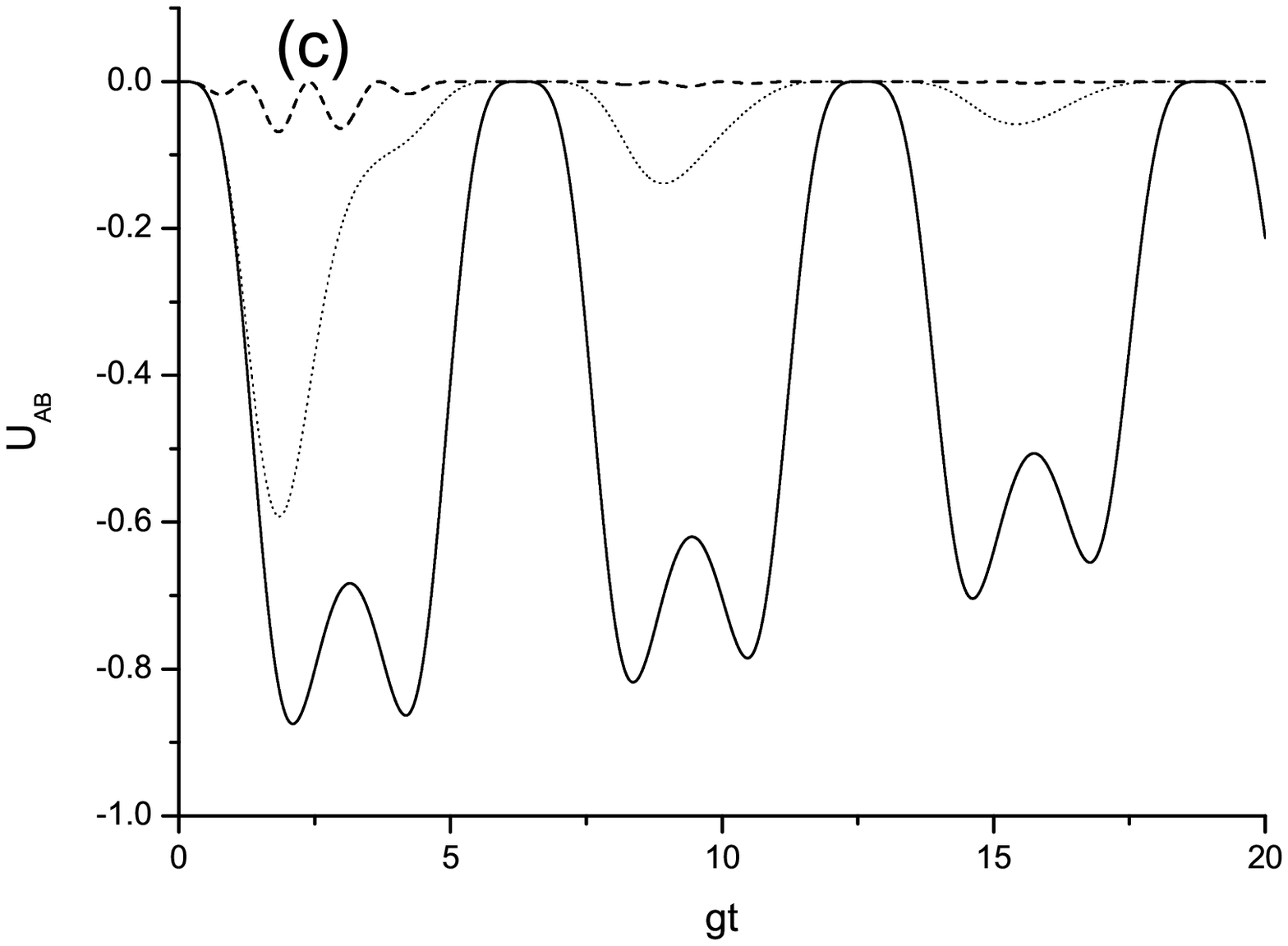}
\vspace*{.2cm}
\caption{\label{switchfunc}
The evolution of  $C_{AB}$,  $P_{AB}$ and  $U_{AB}$ for
different detuning, $\Delta=0.1$(solid), $\Delta=1$(dotted) and
$\Delta=5$(dashed). The two subsystems are symmetric for $k=l=0.1$
and the field-mode structure parameter $p=1$. (a) $C_{AB}$ versus
time $gt$;  (b) $P_{AB}$ versus time $gt$; (c)  $U_{AB}$ versus
time $gt$.}
\end{center}
\end{figure}
%--------%
%-------------------------------------------------------------------
%--------%
%
\section{Entanglement-purity-energy diagram}
In this section, we devote to investigate the relationships among entanglement, purity and energy for atomic subsystem,
which  reflects much of the nontrivial information about the particular atomic state in the atom-field interacting process.
Here, we  limit our study to the  weak thermal field with $k=l=0.1$ and
the field-mode structure parameter $p=1$ in resonant situation.
Under these conditions, $\cos2\theta_{n}=\cos2\phi_{m}=0$, $\sin2\theta_{n}=\sin2\phi_{m}=-1$.
We plot   entanglement-purity-energy diagram in Fig.4(a) and
show its projections on  entanglement-energy and  entanglement-purity planes in Fig.4(b) and Fig.4(c), respectively.
At initial time, $C_{AB}=1$, $P_{AB}=1$ and $U_{AB}=0$,  the atomic qubits in the maximal entangled  state.
From Fig.4(b) we can find that the disentanglement process accompanies by  excitations transferring from atomic subsystem to cavity field modes
and atomic state from a pure state convert to  mixed states.
The  minimal energy for atomic subsystem is about -0.7  when the two atoms are separable,
and the maximum value is zero when the two atoms are in
the maximal entanglement  state.
This suggests the atomic state can not evolve to   $\rho_{AB}=|gg\rangle\langle gg|$ ($U_{AB}=-1$)
or $\rho_{AB}=|ee\rangle\langle ee|$ ($U_{AB}=1$)  in the atom-field interaction process,
which can also be confirmed in Fig.4(c).
When $C_{AB}=0$, $P_{AB}\neq1$,  the atomic qubits is  in mixed state  when they are separable.
While $C_{AB}=1$ corresponding to $P_{AB}=1$, this indicates that the atomic qubits
can just realize the maximal entanglement pure state but the maximal entanglement mixed state can not be obtained.
%%
%-------------------------------------------------------------------
\begin{figure}
\begin{center}
\includegraphics[height=3cm,angle=0]{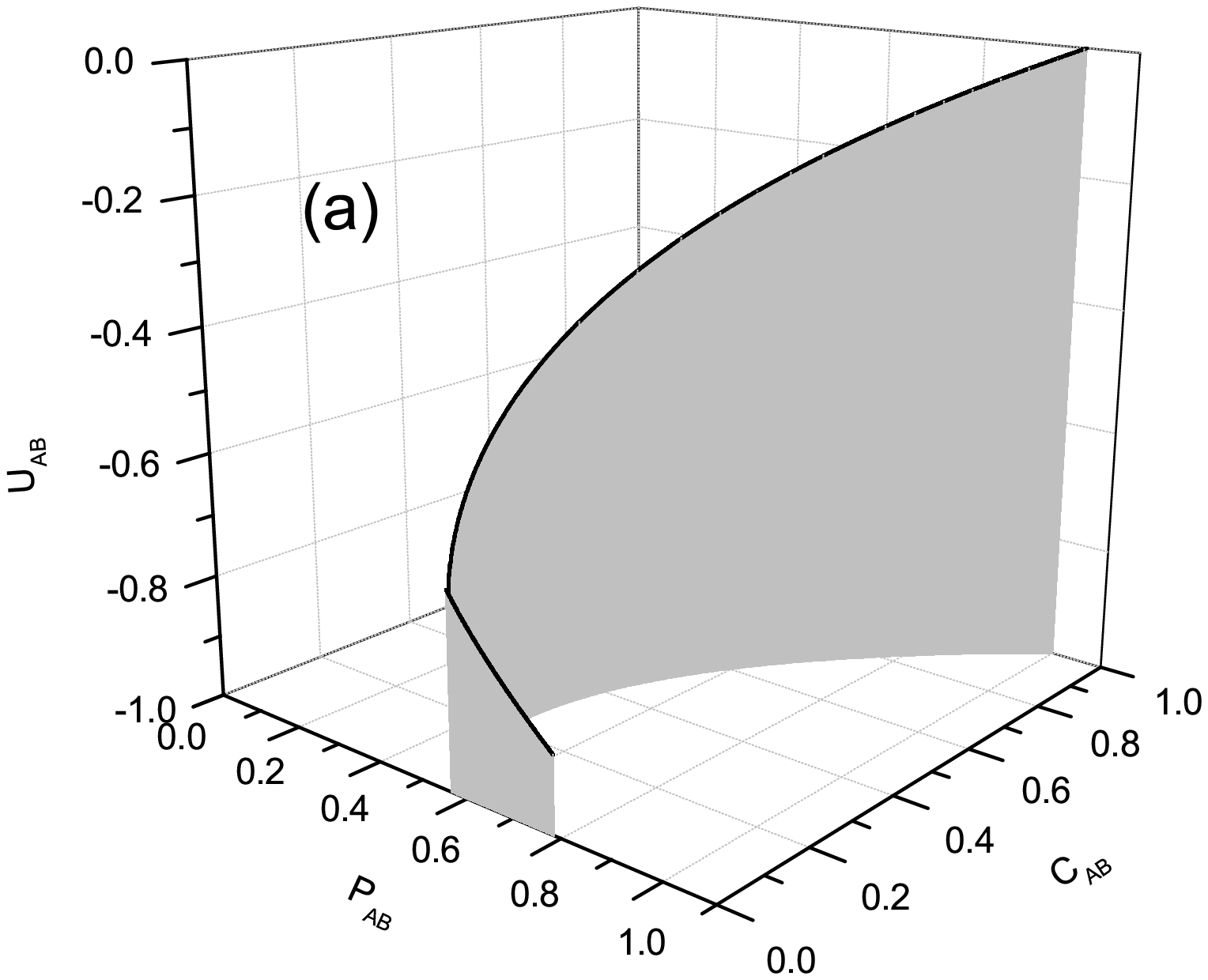}
\includegraphics[height=3cm,angle=0]{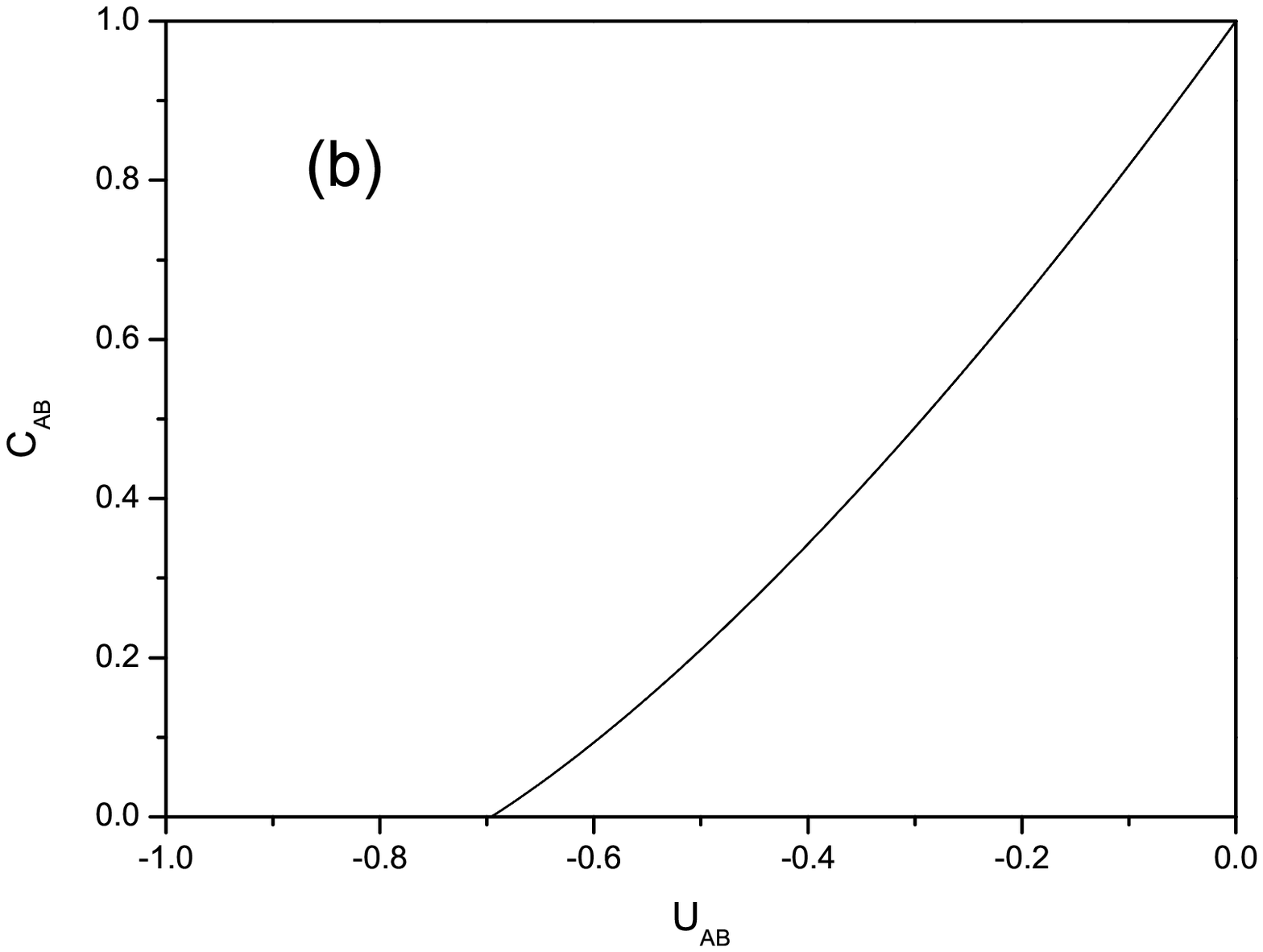}
\includegraphics[height=3cm,angle=0]{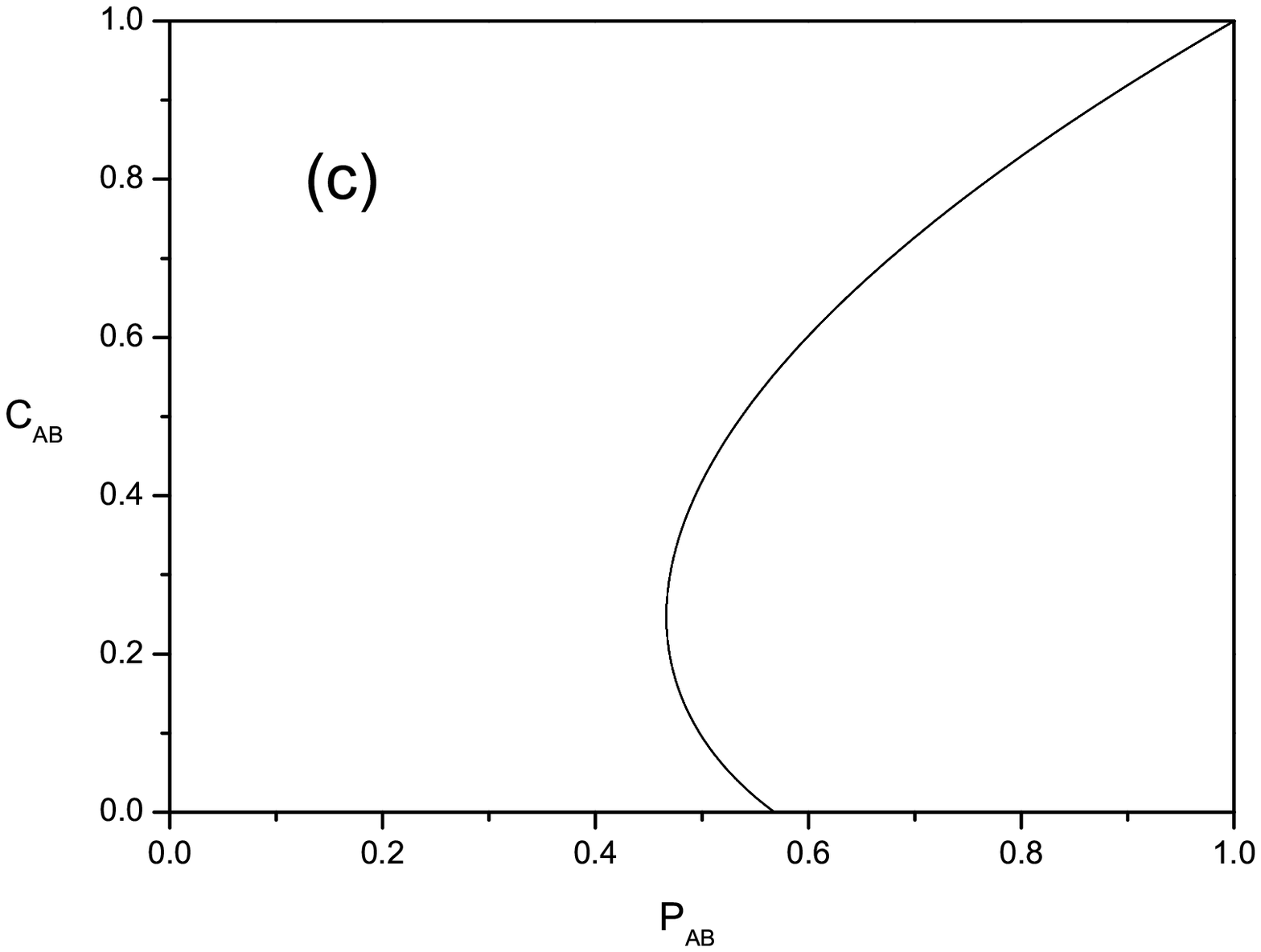}
\vspace*{.2cm}
\caption{\label{switchfunc}
The relationship  between entanglement, mixedness and energy for $p=1$, $\Delta=0$, $k=l=0.1$.
(a)three-dimensional diagram for $C_{AB}$, $P_{AB}$ and $U_{AB}$; (b) $C_{AB}$ versus $U_{AB}$; (c)  $P_{AB}$ versus $J_{AB}$.}
\end{center}
\end{figure}
%--------%
%-------------------------------------------------------------------
%--------
%
%
\section{Conclusion}
In this paper, we employed  three parameters---entanglement, purity and energy to describe the information about
two distant atoms which are initially prepared in Bell state. Our results showed that
considering  the atomic motion and the field-mode structure
can lead to the  periodic evolution of entanglement, purity and energy. With the increase of
field-mode structure parameter $p$, both their evolution periods  and their amplitudes are decreased
while their maximum values are unchanged.
However,  strong thermal field  can reduce the peak values  of entanglement,  purity and energy of the atomic qubits
and make the atomic state initially in a pure state to  mixed states.
Meanwhile, in  such a chaotic field, energy transfer between atoms and fields is  more and more less with the
increase of  thermal field strength.
In addition, large detuning is in favour of reducing their oscillation time  and  "frozing" the initial
maximal entanglement state in the atomic subsystem.
We also analyzed the possible state that the atomic qubits may  evolves into.
From the entanglement-purity-entanglement diagram we found that the  disentanglement process for the atomic subsystem
accompanies both by the  excitation  transferring  from atomic subsystem to cavity field modes
and the state converts  from a pure state  to  mixed states.
Our number results showed that, when the atomic state is in the maximal entanglement  state, it is in a pure state at the same time;
when the two atoms are  separable, the atomic state is in a mixed state. However,
in the atom-field interacting process,  the state for atomic qubits can not evolve to
the maximal entanglement mixed  state.
\section*{Acknowledgment}
This research is supported by the National Natural
Science Foundation of China under Grant No. 10704031, the National
Science Foundation for Fostering Talents in Basic Research of the
National Natural Science Foundation of China Under Grant No.
J0630313, the fundamental Research Fund for Physical and
Mathematical of Lanzhou University Under Grant No. Lzu05001, and the
Natural Science Foundation of Gansu Under Grant No. 3ZS061-A25-035.

\end{document}